\documentclass{article}

\usepackage{booktabs}
\usepackage{graphics}
\usepackage{graphicx}
\usepackage{listings}
\usepackage{nameref}
\usepackage{algorithm}
\usepackage{algpseudocode}
\algrenewcommand\algorithmicrequire{\textbf{Input:}}
\algrenewcommand\algorithmicensure{\textbf{Output:}}
\usepackage{bm}
\usepackage{amsmath}
\usepackage{amssymb}
\usepackage{caption} 
\usepackage{multirow}

\newcommand{\pro}{ProDOMA}
\usepackage{arxiv}

\usepackage[utf8]{inputenc} 
\usepackage[T1]{fontenc}    
\usepackage{hyperref}       
\usepackage{url}            
\usepackage{booktabs}       
\usepackage{amsfonts}       
\usepackage{nicefrac}       
\usepackage{microtype}      
\usepackage{lipsum}
\usepackage{graphicx}
\graphicspath{ {./images/} }

\title{ProDOMA: improve PROtein DOMAin classification for third-generation sequencing reads using deep learning}

\author{
 Nan Du \\
  Dept. of Computer Science and Engineering\\
  Michigan State University\\
  East Lansing, MI 48824, United States\\
  \texttt{dunan00001@gmail.com} \\
   \And
 Jiayu Shang \\
  Dept. of Electrical Engineering\\
  City University of Hong Kong\\
  Kowloon, Hong Kong SAR, China\\
  \texttt{jyshang2-c@my.cityu.edu.hk} \\
  \And
 Yanni Sun \\
  Dept. of Electrical Engineering\\
  City University of Hong Kong\\
  Kowloon, Hong Kong SAR, China\\
  \texttt{yannisun@cityu.edu.hk} \\
}

\begin{document}
\maketitle
\begin{abstract}
\textbf{Motivation:} With the development of third-generation sequencing technologies, people are able to obtain DNA sequences with lengths from 10s to 100s of kb. These long reads allow protein domain annotation without assembly, thus can produce important insights into the biological functions of the underlying data. However, the high error rate in third-generation sequencing data raises a new challenge to established domain analysis pipelines. The state-of-the-art methods are not optimized for noisy reads and have shown unsatisfactory accuracy of domain classification in third-generation sequencing data.  New computational methods are still needed to improve the performance of domain prediction in long noisy reads.\\
\textbf{Results:} In this work, we introduce \pro, a deep learning model that conducts domain classification for third-generation sequencing reads. It uses deep neural networks with 3-frame translation encoding to learn conserved features from partially correct translations. In addition, we formulate our problem as an open-set problem and thus our model can reject unrelated DNA reads such as those from noncoding regions.  In the experiments on simulated reads of protein coding sequences and real reads from the human genome, our model outperforms HMMER and DeepFam on protein domain classification. 
In summary, \pro\ is a useful end-to-end protein domain analysis tool for long noisy reads without relying on error correction. 
\\
\textbf{Availability:} The source code and the trained model are freely available at https://github.com/strideradu/ProDOMA.\\
\textbf{Contact:} \href{yannisun@cityu.edu.hk}{yannisun@cityu.edu.hk}\\
\end{abstract}


\section{Introduction}

\label{sec_background}

Third-generation sequencing technologies, such as Pacific Biosciences single-molecule real-time sequencing (PacBio) and Oxford Nanopore sequencing (Nanopore), produce longer reads than next generation sequencing (NGS) technologies. With increased read length, long reads can contain complete genes or protein domains, making gene-centric functional analysis for high throughput sequencing data more applicable \cite{zhang2019deepfunc, Le19, hong2020protein}. In gene-centric analysis, often there are specific sets of genes in pathways that are of special interest, for example G protein-coupled receptor (GPCR) genes in intracellular signaling pathways for environmental sensing, while other genes in the assemblies provide little insight to the specific questions.

One basic step in gene-centric analysis is to assign sequences into different functional categories, such as families of protein domains (or domains for short), which are independent folding and functional units in a majority of annotated protein sequences. There are a number of tools available for protein domain annotation. They can be roughly divided into two groups depending on how they utilize the available protein domain sequences. One group of methods rely on alignments against the references. HMMER is the state-of-the-art profile search tool based on profile hidden Markov models (pHMM) \cite{eddy1998phmm, el2018pfam}. But the speed of the pHMM homology search suffers from the increase of the  families. Extensive research has been conducted to improve the efficiency of the profile homology search \cite{hmmer}. 

The other group of tools are alignment-free  \cite{davies2007GPCR}. Recent developments in deep learning have led to alignment-free approaches with automatic feature extraction \cite{li2017lstm, seo2018deepfam, Colwell19, kabuka20}. A review of some available methods and their applications can be found in \cite{Karl19}. Of the learning-based tools, the most relevant one to protein domain annotation is DeepFam  \cite{seo2018deepfam}, which used convolutional neural networks (CNN) to classify protein sequences into protein/domain families. 
The authors showed that it outperformed HMMER and previous alignment-free methods on protein domain classification. Also, DeepFam is fast and the speed is not affected much by the number of families. For example, DeepFam is at least ten times faster than HMMER when 1,000 query sequences are searched against thousands of protein families \cite{seo2018deepfam}. Thus deep learning-based methods have advantages for applications that do not need detailed alignments.

Despite the success of existing protein domain annotation tools, they are not ideal choices for domain identification in error-prone reads. In particular, most of these errors are insertions and deletions, which can cause frameshifts during translation \cite{fu2019comparative}. Without knowing the errors and their positions, the frameshifts can lead to only short or non-significant alignments \cite{du2016framepro}. As the translation of each reading frame is partially correct, it also leads to poor classification performance for existing learning-based models.

\subsection{Domain classification with error correction}
As there are error correction tools for third-generation sequencing data \cite{lima2020comparative, fu2019comparative}, an alternative pipeline is to apply tools such as HMMER and DeepFam to error-corrected sequences. 
Error correction tools can be generally divided into hybrid and standalone depending on whether they need short reads for error correction. Recently, several groups conducted comprehensive review and comparison of existing error correction tools \cite{lima2020comparative, fu2019comparative}. None of these tool can achieve optimal performance across all tested data sets.
Particularly, the performance of standalone tools is profoundly affected by the coverage of the aligned sequences against the chosen backbone sequences. When the coverage is low (e.g. <50X for LoRMA \cite{lorma17}), less regions of the reads can be corrected. In addition, we found that error correction tools have difficulty correcting mixed reads from homologous genes within the same family, such as GPCR. The similarities between different genes/domain sequences can confuse the error correction method. 

\subsection{Overview of our work}
\label{subsec_overview}

In this work, we designed and implemented \pro, a deep learning based method to predict the protein domains for third-generation sequencing reads. By training a CNN-based model using 3-frame translation encoding, \pro\ is able to classify error-containing reads into their correct domains with significantly better accuracy. Compared to previous works, \pro\ has several merits. First, it does not require error correction. As a result, it has robust performance for low coverage data. 
Second, although we use simulated PacBio reads as the training data, the experimental results show that \pro\ performs well on real PacBio and Nanopore data. Third, unlike previous deep learning works that were designed for classification, \pro\ can also be used for detection by distinguishing targeted domain homologs from irrelevant coding or non-coding sequences. The detection performance is better than HMMER after \pro\ adopts a modified loss function from targeted image detection. 

The classification accuracy of \pro\ consistently outperformed the state-of-the-art method like HMMER and DeepFam across various error rates. And its performance is not affected by coverage. We tested it on real third-generation sequencing datasets, focusing on its function on detecting targeted domains using Outlier Exposure \cite{hendrycks2018outlierexposure}. \pro\ achieved higher recall with comparable precision to HMMER.

\section{Methods}
\label{sec_methods}

\begin{figure}[htbp!]
    \centering
    \includegraphics[width=0.9\linewidth]{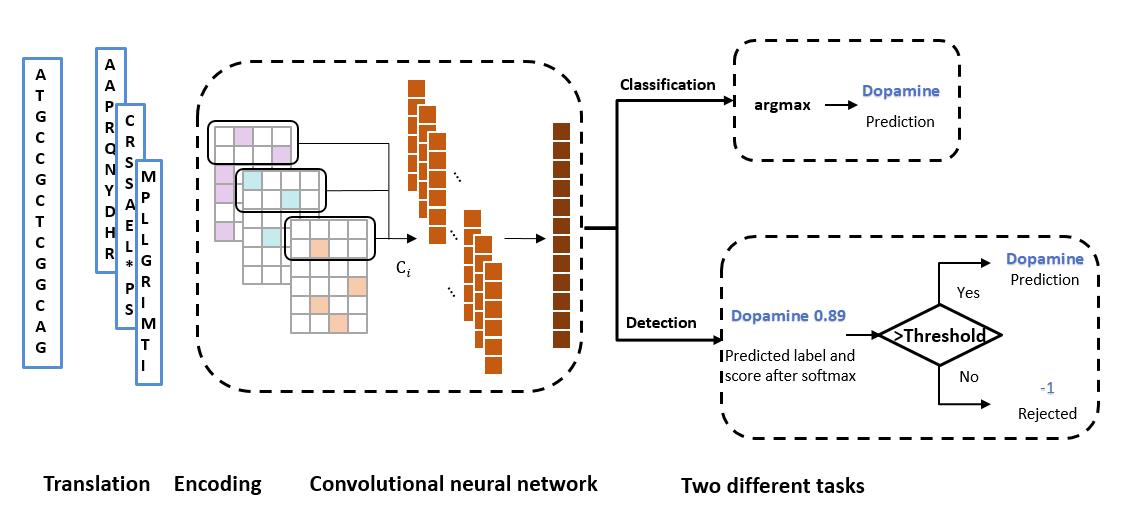}
    \caption{The overview of \pro. The input sequence was translated and encoded to a 3-channel tensor.   $c_i$ is  defined in Equation (1).  In the classification task, the model directly outputs the family with the largest score as the prediction. In the detection task, the maximum of softmax score needs to compare with a specified threshold to determine whether the input contains a trained domain family or should be rejected. }
    \label{fig:tmp}    
\end{figure}

Figure \ref{fig:tmp} sketches the architecture of \pro. It is based on CNN because the convolutional filters can represent motifs, which are important for sequence classification \cite{seo2018deepfam}. 
It incorporates 3-frame encoding to convert DNA reads into a 3-channel tensor as the input to a neural network. From the input, \pro\ automatically extracts features by using convolutional layer and max-over-time pooling layer. A classifier with two fully connected layers was used to generate the probabilities of the sequence against all input protein domains. To exclude the unrelated coding or non-coding DNA sequences, we trained the CNN using a modified loss function so that out-of-distribution samples tend to have uniform distribution on softmax values.

\subsection{Encoding}
 With frequent insertion and deletion errors, the correct translation of a read is composed of fragments of different reading frames. In order to train the CNN to learn the conserved motifs from erroneous translations,  we implemented and compared multiple encoding methods (see Section \ref{sec:otherencoding}). The empirical results show that 
 the 3-frame encoding scheme achieved the best performance. In this scheme, each DNA sequences is translated into 3 protein sequences using 3 reading frames. To accommodate domain identification on the reverse strand, the reverse complement of the sequence can be used as input.  
All the residues in the translated sequence are one-hot encoded using a 21-dimensional vector following IUPAC amino acid code notation. Then we combine three matrices into a single 3-channel tensor like an RGB image. 

Given a translated sequence of length $n$, the encoded input is a tensor with size $3 \times n \times 21$. The pseudo-code of 3-frame encoding can be found in Algorithm~\ref{3frame}. 

\begin{algorithm}
        \caption{3-frame encoding}\label{3frame}
        
        \begin{algorithmic}[1]
        \Require DNA sequence $x$ with length $L$, peptide to index dictionary $idx$, peptide alphabet size $|\Sigma|$, output sequence length $n$.
        \Ensure Input tensor for neural networks with size $3 \times n \times 21$.
            
            \State Initialize an array $arr$ with dimensions $3 \times n \times |\Sigma|$ with all 0
            \For { $i = 1$ to $3$}
            \State $ x^i $ $\gets$ $ x[i:] $
            \State $ y^i $ $\gets$ translation of $ x^i $
            \Comment translate  $ x^i $ into $ y^i$ 
            \For {residue $a$ at position $k$ in $ y^i $}
            \If{$ k $ $ \leq $ $ n $}
            \State $ arr[i, k, idx[a]] $ $\gets$ $ 1 $
            \Comment one-hot encoding for each frame
            \EndIf
            \EndFor
            \EndFor
            \State $ arr $ is the input tensor for neural networks
        \end{algorithmic}
    \end{algorithm}

\subsection{Convolutional Neural Networks}
\pro\ consists of two convolutional layers, one max-over-time pooling layer, one hidden linear layer, and one linear output layer with the softmax function.
For a multi-channel input that we have from 3-frame encoding, we transform the output $arr$ from Algorithm~\ref{3frame} into a feature value using the following equation. 

\begin{equation}\label{conv_multiple}
c_i = f(\sum_{j=1}^{3} \mathbf{w}_{j}\cdot \mathbf{arr}[j][i:i+h-1][1:|\Sigma|] + b )
\end{equation}
$b$ is the bias term and $h$ is the filter size. $f$ is the activation function ReLU \cite{nair2010relu}. The filter consists of three 2D matrices $\mathbf{w}_{j}$ for $j=1,2,$ and $3$, corresponding to three reading frames. $\mathbf{arr}[j][i:i+h-1][1:|\Sigma|]$ defines a 2D window of size $h \times |\Sigma|$ for the one-hot matrix with reading frame $j$.
We applied filters repeatedly to each possible window of the input one-hot matrix to produce the feature map. Then the max-over-time pooling is applied to the feature map to capture the maximum value $\max(c_i)$ as the feature from this particular filter. The max-over-time pooling is flexible with different input length. 

Algorithm~\ref{3frame} described how a single filter in the convolutional layer works. In our application, we used multiple filters with various sizes to extract features of different lengths.

\pro\ has two convolutional layers. The first convolutional layer uses consistent filter size to extract low-level motif-like patterns directly from 3-frame encoding input. Then second convolutional layer extracts high-level, intricate patterns with varying distance from the output of the first convolutional layer. By repeatedly applying the operations, we can finally generate a feature map. Then the max-over-time pooling was applied to keep the most important features. Dropout \cite{srivastava2014dropout} is also used after pooling to prevent overfitting and to learn robust features. A two-layer classifier with softmax function transfers the features to a vector of probabilities over each label. For classification, we select the label with the maximum probability as the prediction from \pro.

\subsection{Comparison of encoding methods and model structures}
\label{sec:otherencoding}
We also tested other encoding methods with similar model structure to \pro: (1) DNA one-hot encoding, which directly transfers DNA sequence to one hot encoding matrix of size  $L \times 4$. For a fair comparison, we used filter sizes that are 3 times as long as we used for 3-frame encoding; (2) 3-branch model, where we constructed a network architecture with three branches processing each of the 3-frame translated protein sequence separately. Each of the branches consists of identical convolution layers, and all the parameters are shared between the same layer of 3 branches. In other words, the Equation (\ref{conv_multiple}) becomes 
$c_i(j) = f( \mathbf{w}_{j}\cdot\mathbf{arr}[j][i:i+h-1][1:|\Sigma|]+ b )$
for $j = 1$ to $3$. In the 3-branch model, each branch models the translated protein sequences separately before the merging layer right before the two-layer classifier. In contrast, in our 3-frame encoding, all three translated protein sequences were processed and combined by the 3-channel convolution filter in the first convolutional layer. 

Our experimental results show that 3-frame encoding is a better encoding scheme, possibly because it can effectively encode the original DNA sequence information and also helps convolutional filters extract useful features for prediction of the protein domains (See results in Section \ref{exp_architecture}). In addition, our experiments show that changing the order of the input reading frames does not affect the classification accuracy. 

\subsection{Detecting out-of-distribution inputs}
We have described how \pro\ predicts the domain labels for given DNA reads using CNN and softmax. However, with the close-set property of softmax, the classifier will always assign a label for the input sample, even if the input is not related to any label in the model (we call such inputs out-of-distribution samples, compared to in-distribution samples). For example, in RNA-Seq data, not every read encodes targeted domain families in the model. In real applications, this close-set property will lead to an undesired high false-positive rate. To address the problem, we adopt Outlier Exposure (OE) \cite{hendrycks2018outlierexposure} with a threshold on softmax prediction probability \cite{hendrycks2016baseline} to distinguish the out-of-distribution inputs from in-distribution ones.

\subsubsection{The threshold baseline}
Usually, the samples from the out-of-distribution dataset tend to have small softmax values because their normalized probabilities are more uniformly distributed than the samples from the in-distribution dataset. 

Following \cite{hendrycks2016baseline}, we extracted the maximum value of the softmax probability vector from the output of \pro\ for each input sample. We separated the in-distribution samples from the out-of-distribution samples by specifying a threshold on the maximum softmax probability. A holdout dataset with both in-distribution and out-of-distribution samples was used to empirically determine the best threshold that can produce the largest $F_1$ score: 
$F_1 = 2 \cdot \frac{\text{precision}\cdot\text{recall}}{\text{recall}+\text{precision}}$.
Then this learned softmax threshold is used to reject any sample with smaller softmax values. The performance of this baseline model is shown in the Figure \ref{fig:softmax_dist}(A).

\subsubsection{Outlier Exposure}

\begin{figure}[htbp!]
    \centering
    \includegraphics[width=1\linewidth]{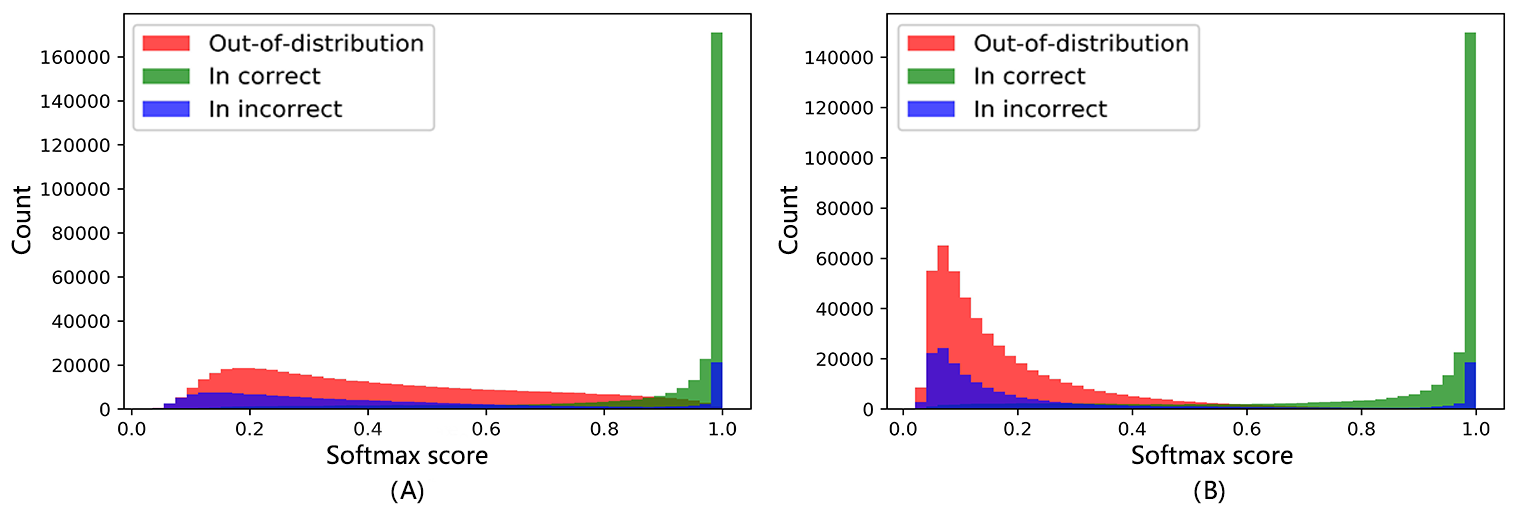}
    \caption{The histograms of maximum softmax values for in-distribution and out-of-distribution samples from base model (\textbf{A}) and model with Outlier Exposure (\textbf{B}). ``In correct'': in-distribution samples with correct classification. ``In Incorrect'': in-distribution samples with incorrect classification. }
    \label{fig:softmax_dist}    
\end{figure}

To further improve the performance of the out-of-distribution sample detection, we adopt the Outlier Exposure (OE) method introduced by \cite{hendrycks2018outlierexposure}. As we discussed previously, we expect the out-of-distribution samples to have uniformly distributed softmax probabilities for all classes. However, as such inputs were never processed in training, sometimes the model will yield unexpected high confidence prediction for out-of-distribution inputs (Figure \ref{fig:softmax_dist}(A)). To address the problem, we expose the model to out-of-distribution samples in the training process to let the model effectively learn the heuristics for detecting out-of-distribution inputs. Compared with the threshold baseline, we need to introduce a new dataset with out-of-distribution samples in the training process.

Given a model $g$ and the learning objective $\mathcal{L}$, the objective of OE is to minimize the original loss function with an auxiliary loss term to regularize the out-of-distribution examples. OE can be formulated as minimizing the following objective \cite{hendrycks2018outlierexposure}:
\begin{equation}\label{oe_objective}
\mathbb{E}_{(x,y)\sim \mathcal{D}_{\text{in}}}\left[\mathcal{L}(g(x),y)+\lambda\mathbb{E}_{x'\sim \mathcal{D}_{\text{out}}}\left[\mathcal{L}_{\text{OE}}(g(x'),g(x),y)\right]\right]
\end{equation}
\begin{equation}\label{oe_loss}
\mathcal{L}_{\text{OE}} = -\sum_{i}\mathcal{P}(i)\ln \frac{\mathcal{Q}(i)}{\mathcal{P}(i)}
\end{equation}


$\mathcal{D}_{\text{in}}$ is the original in-distribution dataset, $\mathcal{D}_{\text{out}}$ is the out-of-distribution dataset for OE. In the original classification task, we use the cross-entropy loss function $\mathcal{L}$. In order to force the out-of-distribution samples to have uniform distribution on all labels, we minimize the KL-divergence between out-of-distribution and the uniform distribution as shown in Equation (\ref{oe_loss}). $\mathcal{P}$(i) is the predicted distribution of out-of-distribution samples from the model and $\mathcal{Q}$(i) is a normalized uniform distribution.
In the experiment, we use $\lambda = 0.5$ for the coefficient of the auxiliary loss. More detailed and comprehensive description of OE can be found in the original publication \cite{hendrycks2018outlierexposure}.

Figure \ref{fig:softmax_dist} presents the distribution of the maximum softmax score for each input sequence with and without OE for the threshold calibration dataset we used in Section \ref{subsec_human}.   Without OE, there are still a lot of out-of-distribution samples with large softmax scores (0.5 to 1). With OE, most of the out-of-distribution samples accumulate with small softmax scores (0 to 0.4). With OE, the overlapping area between the two distributions (red vs combined green and blue) is decreased from 26.06\% to 21.99\%. In addition, for the in-distribution samples with small softmax values, their classification results tend to be wrong (blue in Figure \ref{fig:softmax_dist}). Thus, using OE can provide better classification accuracy at a cost of rejecting some in-distribution samples.

\section{Experiments and results}
\label{sec_results}
To evaluate \pro, we applied \pro\ on both simulated and real datasets: a simulated PacBio G protein-coupled receptor (GPCR) coding sequences (CDS) dataset \cite{davies2007GPCR}, and two real third-generation sequencing datasets of human genome \cite{chaisson2015, pbhuman10x}. GPCR is a large protein family that is involved in many critical physiological processes, such as visual sense, gustatory sense, sense of smell, regulation of immune system activity, and so on \cite{trzaskowski2012GPCR}. 
In addition, GPCR is a very diverse set of protein sequences and thus can pose challenges for classification. It consists of 8,222 protein sequences belonging to 5 families, 38 subfamilies, and 86 sub-subfamilies. Following DeepFam, all the experiments are conducted on the sub-subfamilies. 

We compared the performance of \pro\ with HMMER and DeepFam, which are representatives of alignment-based and alignment-free domain classification tools.  In both experiments, \pro\ was trained with simulated PacBio reads from the GPCR CDS downloaded from NCBI.  The simulation was conducted using a popular simulation tool PBSIM \cite{ono2012pbsim} with default setup and error rates from 1\% to 15\%. Following their instructions and design principle,  HMMER and DeepFam were trained using the correct protein sequences in the GPCR dataset. 

In our first experiment, we tested \pro\ and its alternative implementations on simulated PacBio reads. In the second experiment, we tested \pro\ on real PacBio and Nanopore reads from human data.  
All specific commands, parameters, and output of our experiments can be found along with the source code of \pro.

\subsection{Experiments on simulated PacBio GPCR CDS dataset}
\label{sec_GPCR}

The reference coding sequences of each sub-subfamily are divided into 80\% training samples and 20\% test samples. The number of reference sequences in each class is shown in Table S1 in Supplementary File 1. Then we used PBSIM to generate simulated PacBio reads with 80X coverage on the plus strand for training and test samples with specified error rates. As a result, the training set has 939,888 simulated reads, and the test dataset has 228,388 simulated reads for 86 sub-subfamilies, respectively. Our strategy of conducting simulation after splitting the coding sequences can guarantee that there is no overlap of the GPCR CDS sequences between the training and test datasets, which is important for meaningful evaluations. In our experiments, we used 5-fold cross validation. Thus, the above training and testing dataset construction process was repeated for five times. 

\subsubsection{Performance with different architectures}
\label{exp_architecture}
We conducted a series of experiments by varying the key opponents in our base models: the number of convolutional layers, the number of convolution filters, the size of convolution filters, and different encoding strategies. Totally, we compared 14 different combinations of hyperparameters or architectures in the experiments. 
We listed all variations and their accuracy in Table S2 and Figure S1 in the supplementary file, respectively. 
As shown in Figure S1, the highest accuracy is 86.74\%, which is achieved by using 3-frame encoding with two convolution layers. Based on the comparision, the key factors affecting the performance are the encoding strategies and the size of filters.




\subsubsection{The input order of the reading frames does not change the classification accuracy.}
\begin{figure}[h!]
    \centering
    \includegraphics[width=0.9\linewidth]{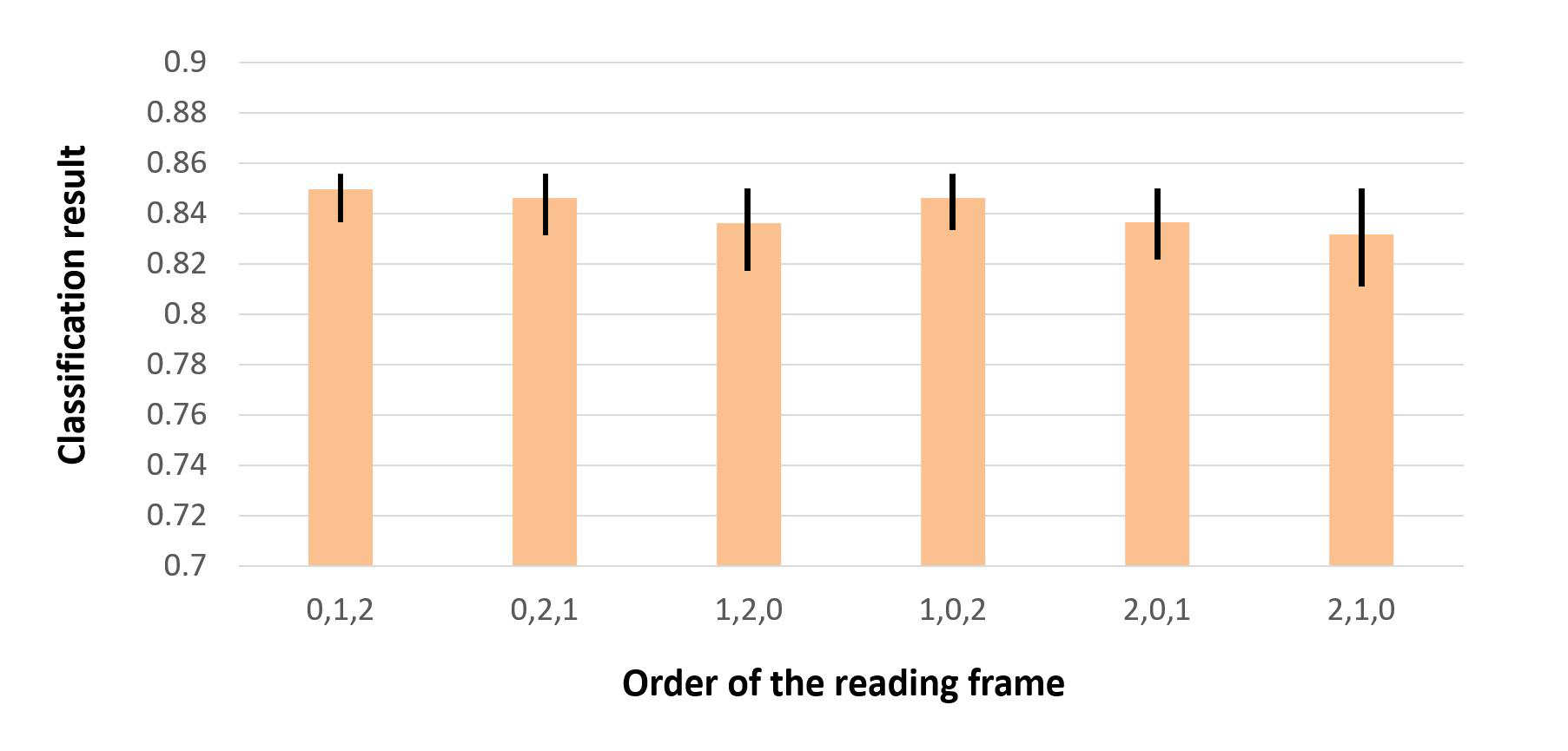}
    \caption{The mean, min, and max value of classification accuracy using different orders of reading frames as input. X-axis: order of reading frames as input. Y-axis: classification accuracy.}
    \label{fig:order}    
\end{figure}

Reads starting from different positions in the same transcript can have different reading frames corresponding to the same translation. In our model training process, the three channels always take translations of reading frame 1, 2, and 3 of a read as input. It is thus fair to ask whether this specific order affects the classification performance. We investigate this question by inputting different orders of reading frames of test sequences to our trained model. Thus, we generated 6 inputs from each reads with different frame orders. As a result, 1,370,328 validation samples are tested in the experiment. Figure \ref{fig:order} shows the classification accuracy of 5-fold cross validation using different reading frame orders as input. The performance keeps nearly the same. 

\subsubsection{Comparison with HMMER and DeepFam}
Following the design principles and the instructions of HMMER and DeepFam, the training of HMMER and DeepFam was conducted using correct protein sequences, rather than DNA sequences. The test sequences are simulated long reads from reference CDs. Their three-frame translations are used as input to HMMER and DeepFam. As long as one of the three translated sequences is classified to the correct sub-subfamily, we call this a correct prediction. 

The classification accuracy of \pro\ and DeepFam was measured using 5-fold cross-validation. 
As it is tedious to perform 5-fold cross validation for HMMER, we used all 5-fold correctly translated protein sequences to train the pHMM model, which will favor HMMER as the trained model has seen the test sequences. MAFFT \cite{katoh2013mafft} was used to generate the multiple sequence alignment for each sub-subfamily. Then we used \lstinline{hmmbuild} in the HMMER package to build pHMM models for each sub-subfamily. For each test DNA sequence, 3-frame translations were applied to get three peptide sequences. All the translated sequences were tested using \lstinline{hmmscan} against all 86 pHMM models we built. 

\begin{figure}[htbp!]
    \centering
    \includegraphics[width=0.8\linewidth]{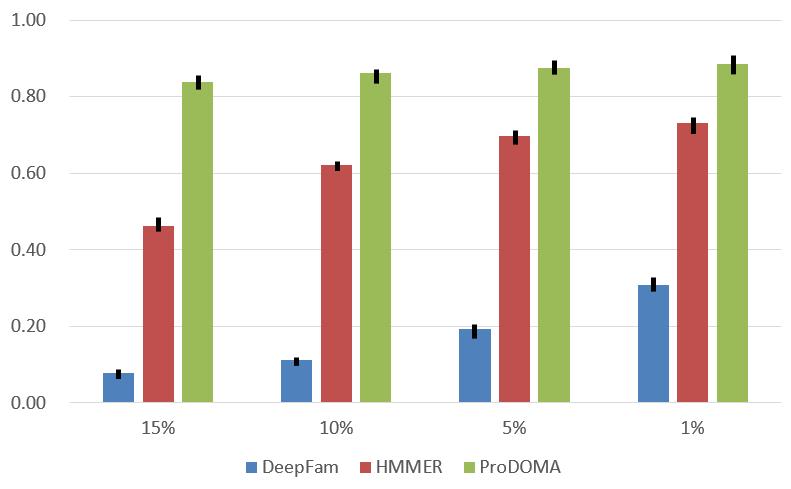}
    \caption{The mean, min, and max value of classification results of \pro, HMMER, and DeepFam on classifying four sets of simulated long reads with different error rates. Each test set contains roughly 228K simulated reads from 86 sub-subfamilies. X-axis: different error rates. Y-axis: classification accuracy.}
    \label{fig:method}    
\end{figure}

Figure \ref{fig:method} compared the classification performance of all methods on the simulated PacBio reads. For this data set, our method achieved better performance for datasets with different error rates. 
The high error rates heavily impacted the performance of HMMER and DeepFam. It is expected because the profile HMM search is much more sensitive to frameshifts caused by gaps, and DeepFam is designed for classifying relatively complete error-free protein sequences.

\subsubsection{Can we rely on error correction?}
As there are error correction tools for long reads \cite{lima2020comparative}, existing domain classification tools such as HMMER can be applied on error-corrected reads. We conducted an experiment to test whether family classification using corrected reads can achieve comparable performance to  classification of error-free sequences. 

Hybrid error correction tools tend to perform better than self-correction tools. However, not every sequence project has budget and manpower to generate both short read and long read libraries and data. Based on these recent reviews  \cite{lima2020comparative, fu2019comparative}, we chose LoRMA \cite{lorma17}, a more recently published self-error correction tool, with the lowest error rate after correction \cite{lima2020comparative}. 
Note that LoRMA failed to generate outputs if we use all the simulated reads as input, probably because of the similarity between the homologous GPCR sequences or the large graph produced by the reads. Thus, we run LoRMA for each set of reads simulated from the same reference sequence of GPCR in order to achieve the best error correction performance. We showed the error correction performance in Table S3 in Supplementary File 1. Although the number of reads remained after error correction increased  with the increase of the coverage, it still discarded a large number of reads (e.g. 75.9\% reads are discarded when the coverage is 30x). Also, the error correction became significantly slower with the increase of the coverage.


\begin{table}[htbp!]
\centering
		\resizebox{\columnwidth}{!}{
            \begin{tabular}{c|c|c|c|c|c|c|c|c}
            \toprule
            \multirow{3}{*}{} & \multicolumn{4}{c|}{All simulated reads}                                         & \multicolumn{4}{c}{Corrected reads} \\ \cline{2-9} & \multirow{2}{*}{ProDOMA} & \multirow{2}{*}{HMMER} & \multicolumn{2}{c|}{DeepFam} & \multirow{2}{*}{ProDOMA} & \multirow{2}{*}{HMMER} & \multicolumn{2}{c}{DeepFam} \\ \cline{4-5} \cline{8-9}
                              & & & Best & \multicolumn{1}{c|}{Average} & & & Best & Average \\ \hline
            10x & 87.12\% & 62.12\% & 25.06\% & 11.07\% & 95.05\% & 97.42\%  & 69.06\% & 32.35\% \\
            20x & 86.47\% & 63.25\% & 25.01\% & 10.98\% & 94.94\% & 97.59\% & 77.36\%  & 37.31\% \\
            30x & 86.69\% & 62.56\% & 25.18\% & 11.02\% & 94.79\% & 98.16\% & 80.76\%  & 38.41\% \\
            \bottomrule
            \end{tabular}
}
\vbox{}	
\caption{The classification accuracy comparison between ProDOMA, HMMER and DeepFam on corrected reads. The first column is the coverage.}
\label{tab:comparision}
\end{table}

We recorded the domain classification results for corrected reads in Table \ref{tab:comparision}. 
HMMER achieved the best accuracy for corrected reads. ProDOMA achieved comparable accuracy on corrected reads. With or without error correction, ProDOMA's performance is consistent with the change of the coverage. Without relying on error correction, ProDOMA can conduct domain classification for all reads and thus lead to a more accurate estimation of the domain/family abundance. 
As DeepFam will assign a family to each translation, it is unknown which one should be chosen in practical applications. ``Best'' means that we regard a read as a correct classification as long as one of the translation is correctly classified. ``Average'' is the performance averaged on three translations. 

\subsubsection{Performance of remote homology search}

\begin{figure}[htbp!]
    \centering
    \includegraphics[width=0.7\linewidth]{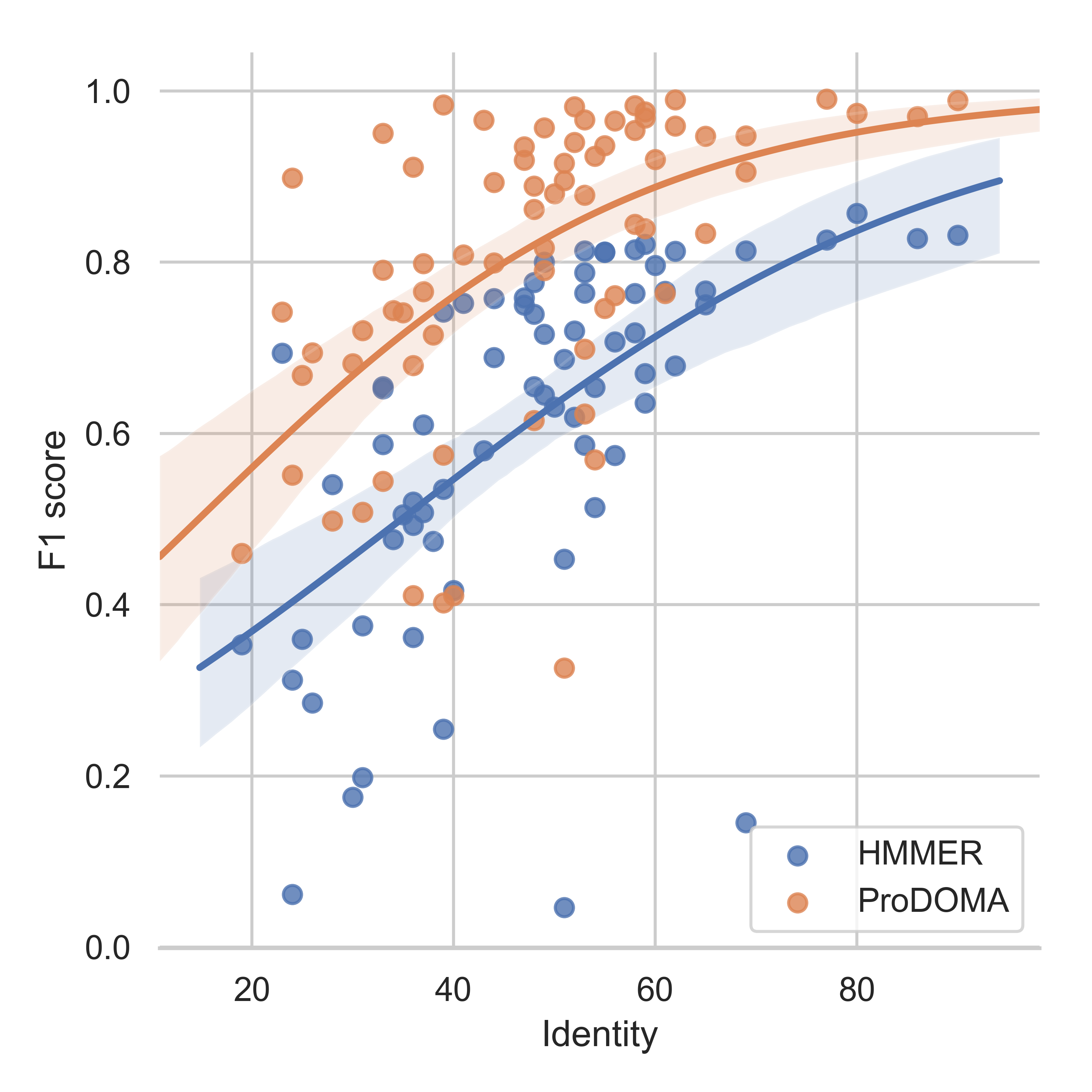}
    \caption{Scatter plot and their logistic regression curves for all sub-subfamilies. The error bar (regions surrounding the curves) is the 95\% confidence interval of the fitting. Y-axis: F1 score of the test sequences. X-axis: intra-class identity. }
    \label{fig:ident_f1}    
\end{figure}

As GPCR is a large protein family, some of their coding sequences can show high diversity, posing a challenge for both alignment-based and alignment-free homology search. We plot the change of F1 score of each protein family with the change of the intra-class identity in Figure \ref{fig:ident_f1}. As DeepFam's performance is inferior to others, we did not include DeepFam in this experiment.  
The intra-class identities were calculated using the \lstinline{alistat} tool provided in the HMMER package. The sub-subfamilies with low identities are more likely to have remote homologs. F1 score is the harmonic mean of recall and precision. For one protein family A, let its true member sequences be set $A^+$ and the predicted sequence set be $A^{pred}$. The recall is thus $ \frac{|{A^+}\cap{A^{pred}}|}{|A^+|}$. Precision is $ \frac{|{A^+}\cap{A^{pred}}|}{|A^{pred}|}$.

Figure \ref{fig:ident_f1} shows that both HMMER and \pro\ suffer from low intra-class identity but the performance of \pro\ deceases slower with the decrease of the identity. One possible reason is that deep learning can learn more degenerate features that are hard to model by HMMs.

\subsubsection{Comparison of the running time}
With a large amount of data generated by third-generation sequencing platforms, we require both high accuracy and efficiency for the algorithms. We run \pro, DeepFam, and HMMER using Intel\textsuperscript{\textregistered} Xeon\textsuperscript{\textregistered} Gold 6148 CPU with 20 cores at the High-Performance Computing Center at Michigan State University. We also tested \pro\ and DeepFam with NVIDIA\textsuperscript{\textregistered} Tesla\textsuperscript{\textregistered} V100 GPU with Apex acceleration library (HMMER doesn't support GPU). For each method, We measured its execution time by averaging 5 independent trials with randomly selected 10,000 sequences.

\begin{table}[htbp!]
	    \centering
		\resizebox{\columnwidth}{!}{
			\begin{tabular}{p{2cm}rrrr}
				\toprule
				Setup & ProDOMA & DeepFam & HMMER & HMMER $\mathtt{-max}$ \\
				\midrule
				CPU   & 1168.78s  & 276.74s & 312.13s & 3470.04s \\
				GPU      & 25.71s  & 20.37s& unavailable & unavailable  \\

				\bottomrule
			\end{tabular} 
		}
		
\vbox{}	
\caption{The average elapsed time to predict sub-subfamily labels of 10,000 simulated PacBio reads for each method. }
\label{tab:time_comapre}
\end{table}

In CPU, HMMER with the default setup runs much faster than \pro\ and DeepFam. One reason is that with high sequencing error rates, the alignment against many candidate sub-subfamilies cannot pass the filter stage of HMMER, skipping the expensive pHMM alignment.
By turning off all filters, the sensitivity of HMMER increases, but at a large cost in speed. With \lstinline{--max} (turning off all filters), HMMER is much slower than deep learning-based methods. With GPU acceleration, the running time of \pro\ is much shorter than the running time of HMMER with the default setup.

\subsection{Human genome dataset}
\label{subsec_human}

To evaluate \pro's performance on real third-generation sequencing dataset, we tested \pro\ on the \textit{H. sapiens} 10x Coverage data from PacBio \cite{pbhuman10x} and Oxford Nanopore Human Reference Datasets Rel6 \cite{jain2018nanopore}. In this experiment, as the real genome sequencing data contains non-GPCR coding sequences, we will also test the performance of \pro\ on detecting out-of-distribution samples.  

\paragraph{Training dataset.}
The training dataset is the same as the previous experiments: long reads simulated using PBSIM.  To apply Outlier Exposure, we constructed a dataset that mixed the previous 5-fold training dataset with an outlier dataset. In order to generate the outlier dataset with similar distributions to the real out-of-distribution samples, we simulated a PacBio human genome dataset from GRCh37/hg19 human reference genome \cite{church2011grch37}. Then we kept the simulated reads that cannot be aligned to any GPCR CDS by BLASR in the outlier dataset \cite{chaisson2012blasr}. As a result, the outlier dataset has 800,000 simulated reads. 
Then we retrained \pro\ with OE as discussed in Methods.

\paragraph{Test dataset.}
We used two test datasets: a PacBio RS II test dataset from PacBio SMRT Sequencing for CHM1TERT human cell line; and a Nanopore test dataset from Oxford Nanopore MinION on CEPH1463 (NA12878/GM12878, Ceph/Utah pedigree).

We determine the ground truth of these test reads using sequence similarity, which favors alignment-based tools.  We first aligned all reads against GPCR CDS dataset using BLASR.  
We extracted the reads with alignment length longer than 60\% of aligned CDS sequence as our in-distribution test samples. For each sample, the ground truth is given by the label of aligned CDS. 

We also randomly selected reads that were not aligned to any GPCR CDS as our out-of-distribution samples. As a result, the dataset consists of 50\% reads that are coding GPCR and 50\% out-of-distribution reads. For reads longer than 3,000 bps in the test dataset, we cropped each read into fragments that are at most 3,000 bps to be consistent with training dataset. As a result, the PacBio test dataset has 12,716 reads, and Nanopore test dataset has 43,654 reads, including both the GPCR CDS and the out-of-distribution samples.


\subsubsection{Out-of-distribution test using PacBio and Nanopore reads}

\setlength{\heavyrulewidth}{1.75pt}
\begin{table}[htbp!]
	    \centering
		\resizebox{\columnwidth}{!}{
			\begin{tabular}{p{3cm}cccccc}
				\toprule
				Method & \multicolumn{3}{c}{PacBio} & \multicolumn{3}{c}{Nanopore} \\
				\midrule
				& Recall & Precision & F1 score & Recall & Precision & F1 score\\
				HMMER   &0.1494 & 0.9507& 0.2583 & 0.3984& \textbf{0.9825}& 0.5670\\
				\pro\ OE & \textbf{0.4479}&\textbf{0.9837}&\textbf{0.6154}&\textbf{0.4836}&0.9731&\textbf{0.6458} \\

				\bottomrule
			\end{tabular} 
		}
		
\vbox{}	
\caption{The performance of protein domain prediction with out-of-distribution examples using  \pro\ with Outlier Exposure (OE), and HMMER on the real PacBio and Nanopore dataset. }
\label{tab:human_detect}
\end{table}

As the purpose of this experiment is to evaluate the performance of GPCR CDS detection, we computed the recall, precision, and F1 score of each tool on labeling reads from the 86 classes. Recall is the ratio of correctly predicted reads to the total number of reads from the 86 classes. Precision is the ratio of the reads from the 86 classes to the total number of reads predicted with the 86 class labels. F1 score is the harmonic mean of recall and precision and we reported micro-F1 score for our multi-classification model.

We benchmarked the OE model with HMMER, which is highly accurate in distinguishing protein domains from other sequences. As shown in Table \ref{tab:human_detect}, both methods have low recall but high specificity because many reads are rejected and not classified into any of the 86 classes. Still, \pro\ with OE achieved significant improvement on recall while the precision is comparable with HMMER (Table \ref{tab:human_detect}). 
In general, both methods have better performance on the Nanopore dataset. Note that in the whole pipeline, we only used simulated PacBio reads for training. This result suggests that our strategy is robust with different types of long reads.

\section{Discussion}
\begin{figure}[h!]
    \centering
    \includegraphics[width=0.65\linewidth]{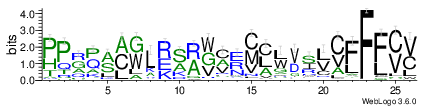}
   \includegraphics[width=0.65\linewidth]{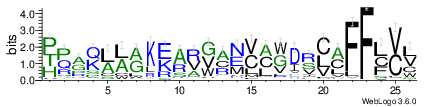}
   \includegraphics[width=0.65\linewidth]{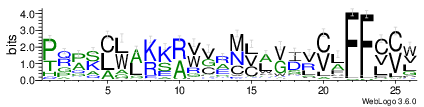}
  \caption{Logos derived from most frequently activated convolutional filters (index: 1033) for Cholecystokinin.}
   \label{fig:cholecystokinin_logo}    

\end{figure}

In this work, we showed that \pro\ can render better accuracy for domain classification in long noisy reads. 
The major differences between the architecture of \pro\ and other CNN-based sequence classification models are the coding strategy designed for erroneous reads and the modified loss function to reject out-of-distribution samples. In order to interpret why this 3-frame encoding works, we extracted sequences corresponding to the most frequently activated filters, which often represent the well-conserved motifs. Then we used Weblogo \cite{crooks2004weblogo} to generate the logos from their alignments as shown in the supplementary file (Figure S3 to Figure S8). An example is provided in Figure \ref{fig:cholecystokinin_logo}. Because we translated the original input DNA sequences into 3 peptide sequences, we have 3 logos associated with 3 reading frames respectively. 
These figures show that the logos corresponding to the three filters are very similar. This is consistent with Figure \ref{fig:order}, revealing that most filters for different channels are learning conserved motifs from error-free regions. 

As the model is trained using sequences with 3,000 nt, padding is applied to all inputs shorter than 3,000. Sequences longer than 3,000 will be cut into substrings of length 3,000, which will then be fed into the model. Thus our CNN model cannot accurately detect the start and end positions of domains.  Instead, a model that can accurately assign weights for each base is needed for detecting the entering and existing positions of each domain. We plan to explore deep learning-based annotation our future work.

In summary, \pro\ provides a complementary tool to current third-generation sequence analysis pipelines on gene-centric function analysis. It can directly identify protein domains in long noisy reads without relying on error correction and its performance is robust to low coverage data and can tolerate higher error rates than other domain classification tools.

\section*{Funding}
The computation was conducted at HPCC of Michigan State University. 
This work has been supported by the City University of Hong Kong 7200620.

\bibliographystyle{unsrt}  
\bibliography{references}  

\end{document}